\documentclass{sig-alternate-per-modified} 

\usepackage{etoolbox}
\usepackage{amssymb}
\usepackage{amsmath}
\usepackage{epsfig}
\usepackage[normalem]{ulem}
\usepackage[linesnumbered,ruled]{algorithm2e}
\usepackage{algorithmicx}
\usepackage{algpseudocode}
\usepackage{setspace}
\usepackage{multirow}
\usepackage{bbm}
\usepackage{epstopdf}
\usepackage{siunitx, booktabs, diagbox}
\usepackage{tikz}
\usepackage{hyperref}
\usepackage{xcolor}
\usepackage{longtable}
\usepackage{balance}
\usepackage{xparse}
\robustify{\cite}

\def\equationautorefname~#1\null{%
  (#1)\null
}

\def \eg {{e.g.,\;}}

\usepackage{enumitem}

\usetikzlibrary{shapes,arrows,positioning}
\usetikzlibrary{calc}
\usetikzlibrary{patterns,decorations.pathreplacing}

\toappear{To appear in \emph{IFIP Performance 2026}, November 2026, Ghent, Belgium.}

\begin{document}

\title{
QPS-ToR: A Parallel Iterative Switching Algorithm for Reconfigurable Optical Datacenter Switching
\thanks{This work was supported in part by the National Science Foundation under Grant No. CNS-2007006 and by a seed gift from Dolby Laboratories.}
}

\numberofauthors{3}

\author{
\alignauthor
Dongzhao Song\\
       \affaddr{Georgia Tech}\\
       \affaddr{Atlanta, Georgia, USA}\\
       \email{dsong84@gatech.edu}
\alignauthor
Qianru Yu\\
       \affaddr{Georgia Tech}\\
       \affaddr{Atlanta, Georgia, USA}\\
       \email{qyu87@gatech.edu}
\alignauthor
Jun ``Jim'' Xu\\
       \affaddr{Georgia Tech}\\
       \affaddr{Atlanta, Georgia, USA}\\
       \email{jx@cc.gatech.edu}
}

\maketitle

\begin{abstract}
Reconfigurable optical data center networks (RODCNs) have emerged as a promising solution for scaling DCN capacity, yet their scheduling mechanisms remain a performance bottleneck: traffic-oblivious schemes inherently limit throughput, while the state-of-the-art traffic-aware scheme, NegotiaToR, uses single-iteration iSLIP as its scheduling engine, which limits throughput to around 60\% and treats all source-destination pairs with equal priority regardless of queue length. We propose QPS-ToR, which replaces NegotiaToR's scheduling logic with SW-QPS, a sliding-window algorithm originally proposed for crossbar scheduling that achieves around 90\% throughput with a single low-complexity iteration. QPS-ToR operates within NegotiaToR's existing workflow, requiring only a revision of the scheduling cycle from three-step Request-Grant-Accept (RGA) to two-step Request-Grant (RG) with a sliding window mechanism.
In flow-level simulations with 128 ToRs under realistic datacenter workloads, QPS-ToR achieves up to 36\% higher throughput and 82\% lower flow completion time (FCT) compared to NegotiaToR, and consistently outperforms RotorNet, a representative traffic-oblivious scheme, on the parallel network topology.
\end{abstract}

\section{Introduction}\label{sec:intro}

The sizes of data center networks (DCNs) and the volumes of network traffic across them continue to grow relentlessly, thanks to existing and emerging data-intensive applications, such as distributed artificial intelligence, cloud computing, and video streaming.  
At the same time, DCN designs are increasingly adopting a flattened (i.e., non-hierarchical) topology, in which Top-of-Rack (ToR) switches are interconnected directly via a flattened fabric of optical switches (\eg in~\cite{sirius2020ballani}) that can be abstracted into a single giant optical DCN switch.
To transport and ``direct'' a massive amount of traffic to their respective destinations, DCN switching solutions capable of connecting many ToR switches, operating at high line rates (\eg \SI{40}{\giga bps}), and delivering high throughput and delay performance, under any admissible traffic demand, are badly needed.

The standard switching solution framework for modern DCN is called RODCN (reconfigurable optical data center network)~\cite{singh2015jupiter, RotorNetDatacenterNetwork, shrivastav-nsdi19-shoal, mellette-nsdi20-opera, sirius2020ballani, Benjamin2020_pulse, urata2022mission, JupiterEvolving, LightwaveFabrics, vamsi-sigmetrics23-mars, zerwas-sigmetrics23-duo, RotorNetDatacenterNetwork, NegotiaToR_2024, SiP-ML-2021-Khani}, in which all ToRs are interconnected via an optical circuit switch (OCS).  Depending on whether or not the OCS configurations during a scheduling epoch vary with the (measured) traffic demand, RODCN schemes can be classified into two types: (1) traffic-oblivious; and (2) traffic-aware.   

\subsection{Traffic-oblivious solutions}\label{subsec:intro_traffic_oblivious}
Traffic switching and forwarding in traffic-oblivious schemes is mostly based on Valiant load-balancing (VLB)~\cite{valiant1982scheme} as follows.  
In each epoch, the OCS follows a fixed, pre-determined sequence of $N$ configurations (matchings), where $N$ is the number of ToR switches in the DCN (This sequence can be partitioned into shorter subsequences when multiple parallel OCS are used, such as in~\cite{RotorNetDatacenterNetwork}).  
These $N$ matchings, by following a cyclic-shift connectivity pattern, allow each input port (server rack) to connect to each output port exactly once during an epoch; we call this all-to-all matchings in the sequel, using the term from~\cite{NegotiaToR_2024}. Here a rack (its ToR switch) functions as both an input port (for receiving traffic from the downlink) and an output port (for sending traffic over the uplink) simultaneously, exactly like an RJ-45 port on an Ethernet switch.   

Now we explain, using an analogy, how traffic is forwarded in such a VLB-based scheme.  These $N$ OCS configurations comprise the 
``daily'' (per scheduling epoch) ``express bus schedule'' 
(evolving connectivities between input-output port pairs).
At each input port (``city''), each ``passenger'' (traffic) will first board the ``first arriving bus'' to the bus' destination say R (that is uniformly random, as intended by VLB), and then from R wait for and board the ``right bus'' to the passenger's destination if it is not R.  
As such, each forwarding decision does not require any nontrivial computation, which allows the traffic-oblivious schemes to be highly scalable (to a large $N$).
For this reason, the vast majority of existing RODCN schemes, such as RotorNet~\cite{RotorNetDatacenterNetwork} that we compare
with in this work, are
traffic-oblivious.  However, the flip side of the coin is that this VLB-based forwarding limits the throughput to around 60\%, since the 
vast majority of passengers need to take two ``bus rides'' to reach their respective destinations.

\subsection{Traffic-aware solutions}
Only three RODCN schemes~\cite{xi-13-petabit, Benjamin2020_pulse, NegotiaToR_2024} belong to the traffic-aware type.
In these traffic-aware schemes, for each epoch, the scheduler needs to compute, based on the (measured) traffic demand, 
a bipartite matching between the input and the output ports that serves as the configuration of the OCS for the epoch.
These matchings need to be of high-quality w.r.t. the traffic demand, in the sense they result in great throughput and latency performances.
As such, this problem formulation is almost identical to that of scheduling a single input-queued (IQ) crossbar (with $N$ input ports and $N$ output ports), 
also known as IQ switching.  Indeed, all three traffic-aware schemes adopt standard IQ switching solutions with adaptations.

When $N$ is large, the standard IQ switching solutions for high-speed routers (e.g., in Juniper's M160 and Cisco's 12000 GSR~\cite{mckeown1997fast}) are the PISA (Parallel Iterative Switching Algorithm) family, partly because a serial matching algorithm is too expensive computationally; for example, the state-of-the-art serial algorithm~\cite{Duan2012MBWP} that computes a maximum weighted (by the traffic demands) matching (MWM), which is considered the highest-quality matching for network switching purposes~\cite{Tassiulas90Max}, has a whopping time complexity of $O(N^3)$~\cite{EdmondsKarp1972}.  
PISA mitigates this high-complexity problem partly by spreading the complexity across the $N$ input ports and the $N$ output ports through parallelism, as follows.
A PISA typically runs multiple iterations, each of which involves typically three back-and-forth rounds of message exchanges between input and output ports, in parallel. These three rounds are commonly called RGA~\cite{Firoozshahian2007CIOQ} in the switching literature:  R round in which each input port sends requests to all output ports it has traffic for;  G round in which each output port grants to one of the requests received;  and A round in which each input port accepts one of the grants received.

In the context of IQ crossbar scheduling, multiple RGA iterations can be completed in a switching cycle, since input and output ports can exchange RGA messages using (extremely short) on-chip wirings between them.  
In the contemporary RODCN context, however, it is impossible to complete multiple RGA iterations in a scheduling epoch, since input and output ports are different ToR switches, and even a roundtrip between an input-output port pair can be longer than an epoch~\cite{NegotiaToR_2024}.  
For this reason, in NegotiaToR~\cite{NegotiaToR_2024}, the state of the art traffic-aware RODCN scheme, only a single RGA iteration of iSLIP~\cite{McKeown99iSLIP} is run to compute a matching (as the OCS schedule), at a performance cost:  when there is no speedup of the ``switching fabric,'' its reported throughput is below 60\%~\cite{NegotiaToR_2024}, which is consistent with the throughput of iSLIP~\cite{McKeown99iSLIP} running a single iteration (after accounting for the OCS reconfiguration delay).
Also for this reason, in NegotiaToR~\cite{NegotiaToR_2024}, even the R, G, and A rounds of an iteration have to spread out to three consecutive epochs, in a pipelined manner.   

\subsection{QPS-ToR Algorithm}\label{subsec:intro_qps_tor}
We were recently aware of SW-QPS (Sliding Window Queue Proportional Sampling)~\cite{meng2021sliding}, a PISA proposed several years ago for IQ crossbar scheduling. SW-QPS can attain high throughput (near or over 90\%
) under all benchmark load matrices, while running only a single low-complexity iteration.
Inspired by this work, in this work we ask and answer the following research question.   
Can we significantly improve the throughput and latency performances of NegotiaToR, by replacing the single-iteration iSLIP with (single-iteration) SW-QPS \cite{meng2021sliding} as its OCS scheduler?  Our answer is affirmative:  with this replacement, the resulting RODCN scheme, which we call QPS-ToR, achieves up to 36\% higher throughput and up to 82\% lower FCT (flow completion time), than NegotiaToR.  

In this work, we make two contributions.  First, we take credit for coming up with this idea and revising the workflow of NegotiaToR to make this replacement happen.  Second, we have resolved a read-after-write hazard caused by the conflict between the sliding-window operation of SW-QPS and the aforementioned pipelined operation of NegotiaToR, at a negligible performance cost.

\section{Background}\label{sec:background}

In~\autoref{subsec:bg_NegotiaToR}, we describe (1) NegotiaToR's epoch structure, which QPS-ToR also follows,
and (2) how its aforementioned pipelined RGA operation fit into the structure. 
In~\autoref{subsec:bg_qps_swqps}, we describe the SW-QPS algorithm, from which QPS-ToR's scheduling engine is adapted.

\subsection{NegotiaToR workflow}\label{subsec:bg_NegotiaToR}

\begin{figure}
  \centering
  \includegraphics[width=0.45\textwidth]{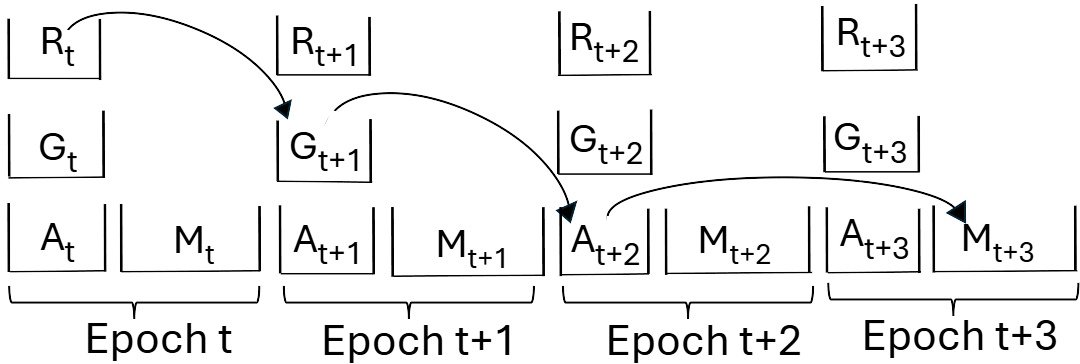}
  \caption{NegotiaToR's Request-Grant-Accept}
  \label{fig:bg_NegotiaToR_RGA}
\end{figure}

In NegotiaToR, time is divided into equal-length intervals called epochs. Figure~\ref{fig:bg_NegotiaToR_RGA} illustrates four consecutive epochs, at times $t$, $t+1$, $t+2$, and $t+3$. Each epoch consists of two phases: a short predefined phase, followed by a longer scheduled phase. During the predefined phases, the OCS supports the aforementioned pipelined (iSLIP's) RGA operation, producing one completed matching per epoch. During each scheduled phase, the resulting matching is applied as the OCS schedule, pairing input and output ports for relatively bulk data transfers.  
As such, the predefined phase serves as an in-band control plane, whereas
the scheduled phase serves as the data plane, of the RODCN.

The pipelined RGA operation, that produces one (relatively) long-duration matching (OCS schedule) per epoch, is carried out as follows.  During each predefined phase (in each epoch), the 
OCS (rapidly) cycles through the aforementioned (in its VLB context in \autoref{subsec:intro_traffic_oblivious}) sequence of $N$ all-to-all matchings
to exchange RGA messages.
As explained earlier, the three rounds (namely R, G, and A) of message exchanges for computing a matching cannot fit in a single epoch and have to span three epochs.  
For example,~\autoref{fig:bg_NegotiaToR_RGA} illustrates an execution path $R_t\rightarrow G_{t+1}\rightarrow A_{t+2}$ that produces (computes) 
a bipartite matching $M_{t+3}$ for use during the scheduled phase of epoch $t+3$.  Here, $R_t$ stands for input ports sending requests to output ports during 
the predefined phase of epoch $t$ according to the measured traffic demand right before time $t$;  and $G_{t+1}$ and $A_{t+2}$ 
stands for the corresponding grants and accepts transmitted during the predefined phases of epochs $t+1$ and $t+2$ respectively.

We can see from this example that without pipelined operation, NegotiaToR can only compute a matching every 3 epochs.  To produce a matching in every epoch, this RGA operation is pipelined, as
shown in
Figure~\ref{fig:bg_NegotiaToR_RGA}.  In each epoch, the predefined phase carries out R, G, A rounds of message exchanges, for three consecutive (in time) matching computations respectively.  For example, $R_t$ for computing
$M_{t+3}$, $G_{t}$ for computing $M_{t+2}$, and $A_{t}$ for computing $M_{t+1}$ are all carried out during the predefined phase at epoch $t$, 

The bulk transfers made by the OCS during the scheduled phases can efficiently handle nearly all large elements (input-output flows) in the traffic demand matrix (TDM).
This efficiency 
comes from the fact that a large TDM element can fill up an edge (input-output connection) in the OCS configuration (matching) during a scheduled phase, which 
effectively amortizes the (nontrivial) OCS reconfiguration delay cost.   
These bulk transfers, however, cannot efficiently handle the mice flows, some of which are also latency-sensitive.  The predefined phases come to the rescue:  the $N$ 
short-duration matchings in each predefined phase give each mice flow, say from input port $i$ to output port $j$, one chance (in the short-duration matching that contains
the edge $(i, j)$) to transmit, by piggybacking after the (pipelined) RGA messages from $i$ to $j$. 

\subsection{QPS and SW-QPS}\label{subsec:bg_qps_swqps}

In this section, we provide a brief introduction to SW-QPS, the state of the art scheduling
algorithm for (the crossbar of) a packet switch, that we will adapt into the scheduling engine of QPS-ToR, our RODCN solution.  
To this end, we first describe QPS (Queue-Proportional Sampling), which SW-QPS builds on. QPS computes a good-quality matching in a single iteration, with both time and communication complexity of $O(1)$~\cite{gong_qps-r_2020}. A QPS iteration~\cite{QPSsampling} contains only two phases: request and grant. In the request phase, each input port sends a pairing request to a single output port with a probability proportional to the length of the corresponding Virtual Output Queue (VOQ);  here a VOQ, say 
VOQ($i$, $j$) is the set of packets queued at the input port $i$ that are destined for the output port $j$. In the grant phase, upon receiving one or more such requests, the output port grants the request accompanied by the largest VOQ length. Unlike iSLIP and most other PISAs (parallel iterative 
switching algorithms), a QPS iteration does not need a separate accept phase, since each input port sends out only one request and hence expects to receive at most one grant.

SW-QPS enhances QPS with a sliding-window (say of length $T$) mechanism that manages, at any time (slot) $t$, the 
computations of $T$ matchings in the sliding-window $(t, t+T]$:  matchings to ``graduate" and be used for 
time slots $t+1$, $t+2$, $\cdots$, $t+T$, which we denote as $M_{t+1}$, $M_{t+2}$, $\cdots$, $M_{t+T}$, 
respectively.  
At time $t$, a single request-grant iteration mostly identical to that in QPS  
is performed to add new edges (input-output pairings) to 
these $T$ matchings.  
The only difference is that each request, say from input port $i$
to output port $j$, contains a bitmap encoding $i$'s availability (whether it has already paired with an output port) during
each of these $T$ time slot; and that, upon receiving this request, $j$ checks its schedule (availabilities in the window), and grants (to $i$) the earliest time slot when both $i$ and $j$ are available.  
Since these request (R) and grant (G) messages contribute to the computations of all $T$ matchings in the window $(t, t + T]$, we denote these two rounds of message exchanges as $R_t^T$ and $G_t^T$ respectively,
with the superscript $T$ emphasizing that this window semantics.

In SW-QPS, after this RG iteration, $M_{t+1}$, the oldest matching in the window, is removed from the window (i.e., graduates) and used as the crossbar configuration for the time $t+1$; and a new empty matching $M_{t+T+1}$ is added to the window.  As such, the window slides one position to the right, becoming $(t+1, t+T+1]$.  From the viewpoint of a matching, say $M_{t+1}$ in this example, it appears in $T$ consecutive sliding windows, namely $(t-T+1, t+1]$, $(t-T+2, t+2]$, $\cdots$, $(t, t+T]$, which provides $M_{t+1}$ with $T$ opportunities (RG iterations) to gain edges before it graduates.

\section{QPS-ToR Algorithm} \label{sec:algorithm}

\begin{figure}[htbp]
  \centering
  \includegraphics[width=0.4\textwidth]{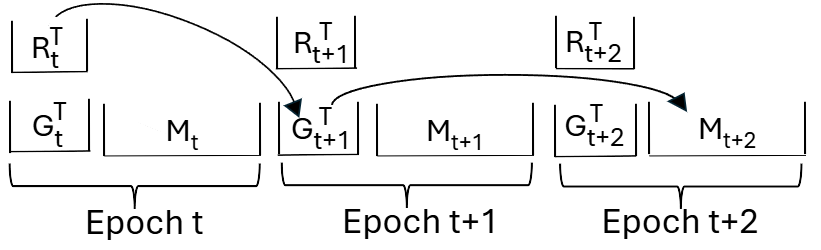}
  \caption{QPS-ToR's Request-Grant}
  \label{fig:QPS-ToR_RA}
\end{figure}

QPS-ToR simply replaces NegotiaToR's pipelined (iSLIP's) RGA operation, illustrated in Figure~\ref{fig:bg_NegotiaToR_RGA}, with a pipelined (SW-QPS') RG operation, illustrated in Figure~\ref{fig:QPS-ToR_RA}.
For example, as shown in Figure~\ref{fig:QPS-ToR_RA}, $R_t^T$ and $G_{t+1}^T$ messages are transmitted in the predefined phases of epochs $t$ and $t+1$ respectively to produce the matching $M_{t+2}$, to be used (as the OCS configuration) for the scheduled phase of epoch $t+2$.  As such, the depth of the pipeline is $2$ here (R and G segments), 
compared to $3$ in NegotiaToR (R, G, and A segments), which translates to further improvement in latency performance beyond that due to SW-QPS being a better (crossbar) switching algorithm than iSLIP.

As mentioned earlier (in \autoref{subsec:intro_qps_tor}), the pipelined RG operation in QPS-ToR has a read-after-write hazard that needs to be fixed. 
We illustrate this hazard by an example.  As shown in~\autoref{fig:QPS-ToR_RA}, the predefined phase contains two segments:  $R_t^T$ and $G_t^T$.
$R_t^T$ messages report the snapshot, right before time $t$, of input ports' availabilities during the sliding-window $(t+1, t+T]$.
In $G_t^T$ messages, output ports add edges to matchings during the sliding-window $(t, t+T-1]$ according to the input ports' availabilities 
reported at time $t-1$ (in $R_{t-1}^T$ messages).  Since $R_t^T$ and $G_t^T$ happen concurrently in the predefined phase of epoch $t$, 
the following ``double booking'' scenario can occur:
$G_t^T$ contain an output port's grant to pair with an input port $i^*$ at epoch $\tau\in (t, t+T-1]$, $i^*$ declares itself available (for pairing) in its request (among
$R_t^T$) to an output port $o^*$ due to this concurrency, and $o^*$ grants (the scheduled phase of) epoch $\tau$ to $i^*$.

QPS-ToR resolves this hazard by running two separate SW-QPS scheduler instances on two disjoint sets of epochs that partition a sliding window, as follows. 
Each request in $R_t^T$ reports the input port's availability at epochs $t+2$, $t+4$, $t+6$, $\cdots$, 
and the grants (in reply to $R_t^T$) in $G_{t+1}^T$ will commit the input ports only to this half of the sliding window.  This partitioning prevents the ``double booking'' in the
example above, because a grant in $G_t^T$ (in reply to $R_{t-1}^T$) makes an input port newly unavailable in an epoch belonging to the other half of the sliding window
($t+1$, $t+3$, $t+5$, $\cdots$).  Although this partitioning in theory reduces an input port's availability ``by half,'' its impact on throughput and latency performance is 
negligible according to our experiments.  Note that NegotiaToR does not suffer from this hazard, because there $R_t$ and $G_t$ messages 
``refer to'' input ports' availabilities during $t+3$ and $t+2$, respectively.

\section{Related Work}\label{sec:related_work}

More than a decade ago, when DCN sizes and link rates were both much smaller than today's, a DCN switching solution framework called hybrid circuit and packet switching was proposed as a cost-effective DCN switching solution~\cite{farrington2010helios, Porter2013TMS, liu2014circuit, Hamedazimi2014FireFly, liu2015scheduling, ProjecToR_2016, bojjacostly, li2017using, vargaftik2016composite, qbvnd-C, 2-HopEclipse, BFF2018, QPSFit2025}.
In addition to an $N\times N$ optical circuit switch (OCS), a hybrid-switched DCN uses an $N\times N$ electronic packet switch (EPS) to connect the $N$ ToRs together.  
The EPS has a lower bandwidth than the OCS, typically by an order of magnitude (\eg \SI{10}{} vs. \SI{100}{\giga bps} per port), but does not incur
a reconfiguration delay (guardband).  The objective of a hybrid-switched DCN solution is to schedule the time-varying configurations (matchings) of its OCS over each epoch for the best possible throughput 
performance.  This scheduling problem has three major differences with that in an RODCN.  First, in hybrid switching, each 
epoch lasts milliseconds.  In comparison, to achieve much lower latency (FCT) demanded by today's latency-sensitive DCN applications, an epoch
lasts only microseconds in an RODCN, like in this work.  Second, hybrid switching solves the following batch scheduling problem: given a traffic demand matrix (TDM)
$\mathcal{D}$ between the ToRs during an epoch, compute a set of configurations (matchings), of varying durations, that OCS should use over time, such that the OCS transmits the bulk of $\mathcal{D}$ and leaves a residue workload that is small enough for the EPS to handle.  In comparison, RODCN computes one fixed-duration (the length of the scheduled phase in an epoch) matching at a time.  Third, a hybrid-switched DCN has an EPS
as a ``dedicated helper'' (e.g., to handle most of mice flows in $\mathcal{D}$) whereas RODCN does not.  

Hybrid switching was an appealing solution because, with an ``ideal separation of duties'' between OCS (mostly for handling elephant flows) 
and EPS (for handling mice flows), the combined (OCS+EPS) switching system can achieve a much higher throughput performance than an OCS alone (\eg using optical switching algorithms such as~\cite{towles2003guaranteed, li2003scheduling, fu2013cost, wang2015end, wang2017heavy}); and this performance gain is much greater than that can be attributed to the minute bandwidth (or contribution) of the EPS, as shown in~\cite{BFF2018}.  
However, when the DCN size becomes larger, all existing hybrid switching algorithms except~\cite{QPSFit2025} become too computationally expensive to scale accordingly.
For example, only QPS-Fit~\cite{QPSFit2025}, the fastest hybrid switching algorithm thanks to its massively parallelable design, can compute an OCS schedule 
for an epoch within an epoch's time (of \SI{3}{\milli\second}) when $N = 96$, whereas all other hybrid switching algorithms require computation times that are 
one to three orders longer.  

This relative (to RODCN) advantage of hybrid switching having an EPS as a dedicated helper became modest only recently,
when OCS with negligibly small guardband (e.g., ranging from \SI{10}{\nano\second} to \SI{100}{\nano\second} in NegotiaToR~\cite{NegotiaToR_2024}) became available at a reasonable cost,
which has moderated the aggregate guardband time spent on the large number of mice flows.  However, without a good scheduling
algorithm, this improved optical hardware capability does not automatically translate into much better FCT and throughput.  Our QPS-ToR is 
such an algorithm.

\section{Evaluation}\label{sec:evaluation}

\subsection{Evaluation setup}

To ensure a fair head-on comparison, we use the following evaluation setup, which is largely identical to that used in the NegotiaToR evaluation~\cite{NegotiaToR_2024}.

\noindent
\textbf{Network setup.} Same as in~\cite{NegotiaToR_2024}, the network consists of 8 parallel optical switches, each implemented as a 128$\times$128 AWGR (arrayed waveguide grating router), interconnecting 128 ToR switches. Each ToR is viewed both as an input port and as an output port, following the RJ-45 analogy in Section~\autoref{subsec:intro_traffic_oblivious}. Each ToR is equipped with 8 optical transceivers, where the $i$-th transceivers across all 128 ToRs are interconnected through the $i$-th AWGR.  

\noindent
\textbf{Epoch settings.} We adopt the same epoch length and structure as in~\cite{NegotiaToR_2024}. During the predefined phase, the 8 optical switches collectively cycle through the aforementioned (in its VLB context in \autoref{subsec:intro_traffic_oblivious}) sequence of $N=128$ predetermined short-duration all-to-all matchings. Since the 128 matchings are distributed across 8 parallel optical switches, the predefined phase consists of only 128/8=16 matching intervals. Each interval lasts \SI{60}{\nano\second}, consisting of a \SI{10}{\nano\second} reconfiguration delay (called the guardband in~\cite{NegotiaToR_2024}) followed by \SI{50}{\nano\second} of data transmission. Consequently, the predefined phase lasts 16*60=\SI{960}{\nano\second}. The scheduled phase is set to \SI{2.7}{\micro\second}, which was found in~\cite{NegotiaToR_2024} to achieve a near-optimal throughput-latency tradeoff for NegotiaToR. Therefore, the total epoch length is 2.7+0.96=\SI{3.66}{\micro\second}.

\noindent
\textbf{Evaluation metrics.} As in~\cite{NegotiaToR_2024}, we focus on two evaluation metrics: (1) latency, measured by average flow completion time (FCT); and (2) throughput.
Also following~\cite{NegotiaToR_2024}, a flow is defined in terms of ToR-to-ToR traffic exchange, i.e., each flow originates at a source ToR and terminates at a destination ToR, and FCT is measured on a per-flow basis.
Preliminary experiments indicate that the improvement trend also holds for 99th-percentile FCT.

\noindent
\textbf{Baselines.}
We compare QPS-ToR against two schedulers:
NegotiaToR~\cite{NegotiaToR_2024, gao2015phost}, the state-of-the-art on-demand traffic-aware RODCN scheduler; and RotorNet~\cite{RotorNetDatacenterNetwork,RealizingRotorNet2024}, the state-of-the-art VLB-based traffic-oblivious RODCN scheduler (described in Section~\autoref{subsec:intro_traffic_oblivious}). All schedulers are implemented within the same \href{https://github.com/NetSys/simulator}{YAPS}~\cite{NegotiaToR_2024,gao2015phost} simulator framework to ensure a controlled comparison. Specifically, the NegotiaToR implementation is obtained directly from its authors, QPS-ToR extends the NegotiaToR codebase with our scheduling logic, and RotorNet is implemented based on the descriptions in~\cite{RotorNetDatacenterNetwork}.
As a VLB-based traffic-oblivious scheduler, RotorNet does not include a scheduled phase. Hence, to ensure a fair comparison, RotorNet uses the same epoch length of \SI{3.66}{\micro\second}, with the entire epoch devoted to the predefined phase. Similar to the other schedulers, the predefined phase is divided into 16 time slots, during which the 8 parallel optical switches collectively cycle through the aforementioned 128 all-to-all matchings. Each slot lasts $\SI{3.66}{\micro\second} / 16 \approx \SI{229}{\nano\second}$, consisting of a \SI{10}{\nano\second} guardband and approximately \SI{219}{\nano\second} of data transmission. This configuration yields a duty cycle of approximately 95.6\% over the rotation.

\noindent
\textbf{Workload characteristics.} Following~\cite{NegotiaToR_2024}, each experiment lasts \SI{30}{\milli\second}. We generate three workloads from the following publicly available DCN traces using the same methodology as in~\cite{NegotiaToR_2024}, where flows arrive according to a Poisson process and their size distribution follows the corresponding trace.
Trace 1 is collected from Meta's Hadoop clusters~\cite{roy2015inside}. This trace is heavy-tailed: 60\% of flows are smaller than \SI{1}{K\byte}, while more than 80\% of the traffic volume comes from elephant flows larger than \SI{100}{K\byte}.
Trace 2 is a more heavily tailed web-search workload~\cite{alizadeh2010data}, in which more than 80\% of flows exceed \SI{10}{K\byte}.
Trace 3 is a lighter-tailed Google datacenter workload~\cite{Homa2018}, where more than 80\% of flows are smaller than \SI{1}{K\byte}.
As in~\cite{NegotiaToR_2024}, each of the 8 optical switches operates at \SI{100}{\giga bps}, collectively providing up to \SI{800}{\giga bps} of aggregate throughput to and from each ToR. We measure network throughput and FCT under offered loads ranging from 10\% to 90\%.   
A key distinction from the original NegotiaToR evaluation lies in how the offered load is normalized.  NegotiaToR reports 100\% throughput in its original evaluation due to a 2$\times$ 
link-rate speedup: while each ToR provides \SI{800}{\giga bps} of uplink capacity, the offered traffic load is normalized to only \SI{400}{\giga bps}. In contrast, our evaluation assumes no link-rate speedup, exposing the full \SI{800}{\giga bps} uplink capacity to offered loads of up to \SI{800}{\giga bps}. Under this setting, no scheduler can achieve 100\% throughput.

\subsection{Evaluation Result}

\begin{figure}[t]
  \centering
  \includegraphics[width=0.5\textwidth]{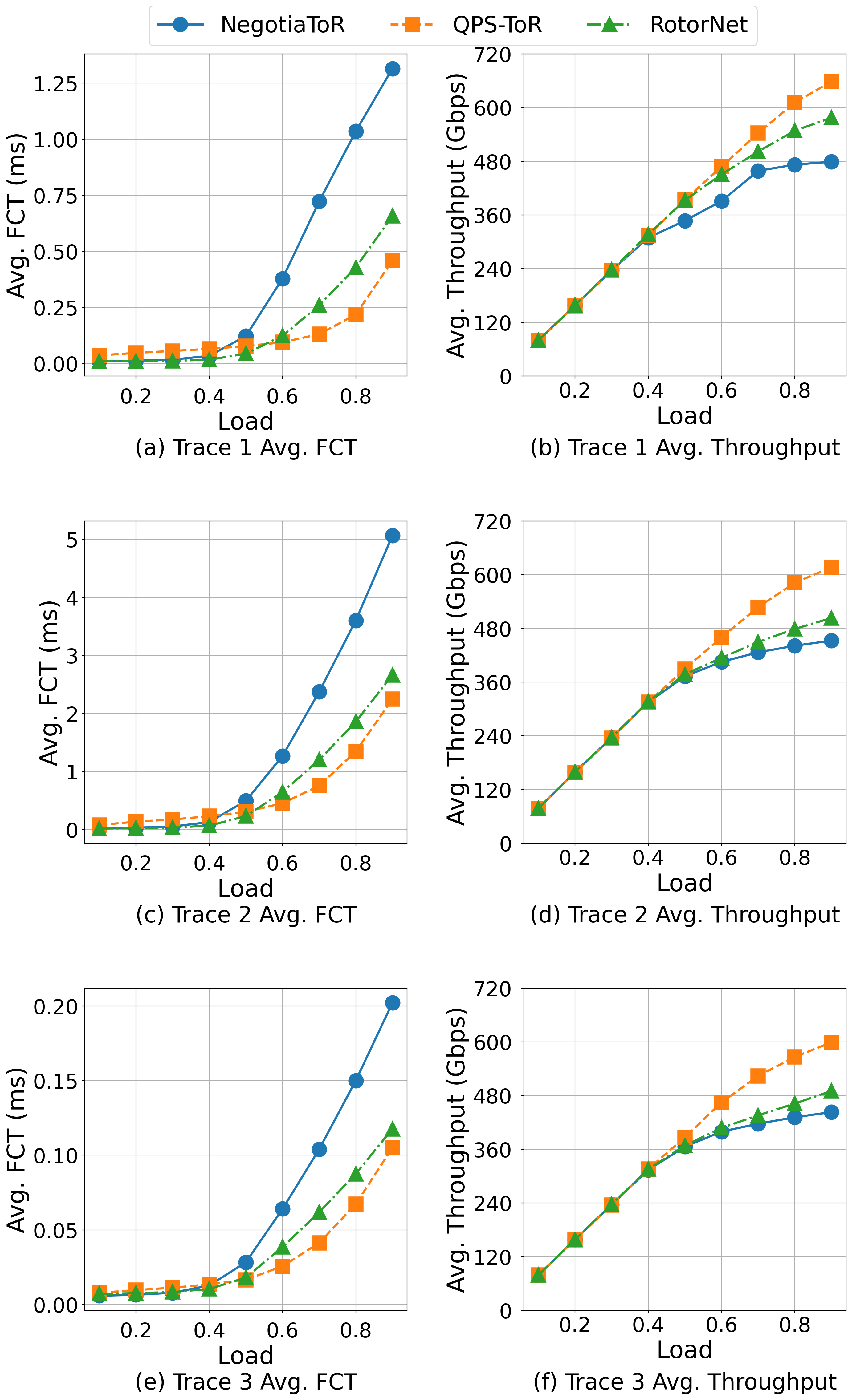}
  \caption{FCT and throughput comparison under three workloads.}
  \label{fig:main_result}
\end{figure}

\noindent
\textbf{Overall FCT.}
As shown in Figure~\ref{fig:main_result}, RotorNet consistently achieves lower FCT than NegotiaToR across all three workloads, suggesting that NegotiaToR's queue-oblivious scheduler fails to fully utilize available optical capacity. QPS-ToR further outperforms both baselines across all workloads on the parallel topology at loads level above
0.5.

Under the default Hadoop workload (Trace~1, Figure~\ref{fig:main_result}(a)), QPS-ToR reduces average FCT by roughly 72\% relative to NegotiaToR and 35\% relative to RotorNet at loads above 0.5. The gap further widens at load 0.7, reaching 82\% and 50\%, respectively.

Under the heavier-tailed web-search workload (Trace~2, Figure~\ref{fig:main_result}(c)), where elephant flows dominate traffic volume, the advantage remains significant: QPS-ToR improves FCT by up to 68\% over NegotiaToR and 37\% over RotorNet at high loads. This behavior is expected, since queue-proportional sampling is most effective when large backlogs provide strong signals for constructing demand-aware matchings.

Under the lighter-tailed Google datacenter workload (Trace~3, Figure~\ref{fig:main_result}(e)), where mice flows dominate, the improvement is more modest---roughly 50\% over NegotiaToR and 21\% over RotorNet---because the reduced prevalence of elephant flows limits the benefit of queue-aware scheduling.

Across all three traces, the improvement arises from the same underlying mechanism: QPS-ToR dynamically steers optical capacity toward heavily loaded ToR pairs, draining large backlogs before substantial queue buildup occurs.

\noindent
\textbf{Overall Throughput.}
As shown in Figures~\ref{fig:main_result}(b)(d)(f), QPS-ToR consistently achieves higher throughput than both baselines across all workloads and load levels.

Under Trace~1 (Figure~\ref{fig:main_result}(b)), QPS-ToR improves throughput by approximately 24\% over NegotiaToR and 8\% over RotorNet at loads above 0.5. Under the heavier-tailed Trace~2 (Figure~\ref{fig:main_result}(d)), throughput gains follow a similar trend: QPS-ToR exceeds NegotiaToR by 22\% and RotorNet by 15\% at loads above 0.5. The gap further widens at load 0.9, reaching 36\% and 22\%, respectively. Under the lighter-tailed Trace~3 (Figure~\ref{fig:main_result}(f)), the throughput advantage is smaller in absolute terms but remains consistent across all loads.

In all cases, NegotiaToR leaves optical capacity underutilized because its scheduling decisions are independent of queue occupancy, whereas QPS-ToR allocates matchings according to observed demand. Even without workload-specific parameter tuning, QPS-ToR demonstrates robust performance advantages across all three traffic distributions.

\section{Conclusion}\label{sec:conclusion}

In this paper, we propose QPS-ToR, a traffic-aware RODCN scheduler that replaces NegotiaToR's single-iteration iSLIP engine with SW-QPS, a sliding-window queue-proportional sampling algorithm adapted from crossbar scheduling.
QPS-ToR operates within NegotiaToR's existing epoch workflow, requiring only a reduction of the scheduling pipeline from three-step RGA to two-step RG, and resolves the resulting read-after-write hazard by interleaving two independent scheduler instances on disjoint epoch subsets.
In flow-level simulations with 128 ToRs under three realistic datacenter workloads, QPS-ToR reduces flow completion time by up to 82\% compared to NegotiaToR and up to 50\% compared to RotorNet, while improving throughput by up to 36\% over NegotiaToR and up to 22\% over RotorNet, consistently outperforming both baselines at loads above 0.5 across all traffic distributions.

{

\balance
\bibliographystyle{abbrv} 

\bibliography{bibs/load_balancing,bibs/QPS,bibs/serenade,bibs/optical_switch,
    bibs/liang-references,bibs/qps-references,bibs/evaplan,bibs/past_pub}
}

\end{document}